\documentclass[11pt,a4paper]{article}

\usepackage{graphicx,eurosym}
\usepackage{float}
\usepackage{afterpage}
\usepackage{epstopdf}
\usepackage{epsfig,cite}
\usepackage{amssymb}
\usepackage{amsmath}
\usepackage{dsfont}
\usepackage{multirow}
\usepackage{url}
\usepackage[colorlinks=false]{hyperref}

\usepackage{url}
\usepackage{hyperref}

\newcommand{\be}{\begin{equation}}
\newcommand{\ee}{\end{equation}}
\newcommand{\bea}{\begin{eqnarray}}
\newcommand{\eea}{\end{eqnarray}}
\newcommand{\bi}{\begin{itemize}}
\newcommand{\ei}{\end{itemize}}
\newcommand{\ben}{\begin{enumerate}}
\newcommand{\een}{\end{enumerate}}

\def\frac#1#2{{{#1}\over {#2}}}
\def\gsim{\mathrel{\rlap{\lower4pt\hbox{\hskip1pt$\sim$}}
    \raise1pt\hbox{$>$}}}         
\def\lsim{\mathrel{\rlap{\lower4pt\hbox{\hskip1pt$\sim$}}
    \raise1pt\hbox{$<$}}}         

\newcommand{\draft}[1]{}

\def\beq{\begin{equation}}  
\def\eeq{\end{equation}}  

\def \n0{N_j^{(0)}}

\def\lapprox{\lower .7ex\hbox{$\;\stackrel{\textstyle <}{\sim}\;$}}
\def\gapprox{\lower .7ex\hbox{$\;\stackrel{\textstyle >}{\sim}\;$}}

\begin{document}

\begin{flushright}
TIF-UNIMI-2015-9\\
\end{flushright}

\vspace{0.4cm}

\begin{center} 
  {\bf{\Large Cost-Benefit Analysis of
 the Large Hadron Collider  to 2025 and beyond}}
\vspace{.7cm}

Massimo Florio$^1$, Stefano Forte$^2$, 
 and Emanuela Sirtori$^3$ 

\vspace{.3cm}
{\it~$^1$ Dipartimento di Economia, Management e Metodi Quantitativi,\\
Universit\`a di Milano , via Conservatorio 7, I-20122 Milano, Italy\\
~$^2$ TIF Lab, Dipartimento di Fisica, Universit\`a di Milano and\\
INFN, Sezione di Milano, Via Celoria 16, I-20133 Milano, Italy\\
~$^5$ CSIL, Centre for Industrial Studies\\
Corso Monforte 15,  I-20122 Milano, Italy}
\end{center}

\vspace{0.1cm}

\begin{center}
{\bf \large Abstract}

\end{center}
{ Social cost-benefit analysis (CBA) of projects has been successfully applied
in different fields such as transport, energy, health, education, and
environment, including climate change. It is often argued that it is 
impossible  to extend the
CBA approach to the evaluation of the social impact of research
infrastructures, because the final
benefit to society of scientific discovery is generally
unpredictable. Here, we propose a quantitative approach to this
problem, we use it to design an empirically testable CBA
model, and we apply it to the the Large Hadron Collider (LHC), the
highest-energy accelerator in the world, currently operating at CERN.
We show that the evaluation of benefits can be made quantitative by
determining their
value to users (scientists, early-stage researchers, firms, visitors)
and  non-users (the general public).  Four classes of
contributions to users are identified:
 knowledge output,
human capital development, technological spillovers, and cultural 
effects. Benefits for non-users 
can be estimated, in analogy to public
goods with no practical use (such as environment preservation), 
using willingness to pay. 
We  determine the probability distribution of
cost and benefits for the LHC  since 1993 until planned decommissioning
in 2025,  and we find there is a 92\% probability that benefits
exceed its costs, with an expected net present value (NPV) of about 3
billion \euro, not including the  unpredictable
economic value of discovery of any
new physics. We argue that 
the evaluation approach proposed here
can be
replicated for any large-scale research infrastructure,
thus helping the decision-making on competing projects, with a
socio-economic appraisal complementary to other evaluation criteria.}

\clearpage


Cost-benefit analysis (CBA) is used to
evaluate the socio-economic impact of any project: it requires~\cite{dreze,johanson,boardman,florio}   the forecasting of inputs, outputs, their marginal social
values (MSV) in order to determine the expected net present
value (NPV) of a project. A project is socially valuable if its
benefits exceed costs over time,  ${\rm NPV}>0$. If $B_{t_i}$ and
$C_{t_i}$ are respectively benefits and costs  incurred at various times 
time $t_i$, 
\begin{equation}\label{eq:npvdef}
{\rm NPV}= \sum_i \frac{B_{t_i}-C_{t_i}}{(1+r)^{t_i}},
\end{equation}
 with $r$ the social
discount rate, needed to convert a future value of $t$ in terms of a reference
$t=0$.  
This
approach is well developed for conventional infrastructures and
supported by the European Commission, the World Bank, the European
Investment Bank, and other national and
international institutions~\cite{baum,EC,EIB,OECD,WHO}. The application of
CBA to 
research infrastructures (RI) has been hindered by  the
unpredictability of future economic benefits of science. 

In order to address the problem quantitatively, 
borrowing ideas from environmental CBA~\cite{pearce,johanson2,atkinson},
we break down the
NPV of a RI in two parts: net use-benefits ${\rm NPV}_u$,
 and the non-use value of the expected discovery $B_n$.
The former, ${\rm NPV}_u$
is the sum of capital and operative cost, and  the
 economic value of its benefits, in turn determined by asking who its
 beneficiaries are. It is an intertemporal value, i.e. it has the
 structure of Eq.~(\ref{eq:npvdef}).
 The latter, $B_n$, captures two types of non-use  values related to
 future discoveries: their quasi-option value (${\rm QOV}_0$)~\cite{arrow}, which
 includes  any future but unpredictable economic benefit of science,
 and an
 existence value related to pure new knowledge per se (${\rm
   EXV}_0$). It is an instant value, i.e. it refers to time $t=0$. 

In order to determine  ${\rm NPV}_u$, we ask
who the beneficiaries of a
 RI are, and thus identify four benefits: publications, to scientists
  (${\rm SC}$), technological externalities, e.g. to firms, 
(${\rm TE}$), human capital formation e.g. for  students and
 postdocs  (${\rm HC}$), and cultural
 effects e.g.  for outreach beneficiaries (${\rm
   CU}$). Costs are determined as the sum  of the economic value of
 capital  (${\rm K}$), labour cost of scientists  (${\rm LS}$) and other
  staff  (${\rm LO}$), and 
 operating costs  (${\rm O}$).

\begin{figure}[t]
\begin{center}
  \includegraphics[width=0.6\textwidth]{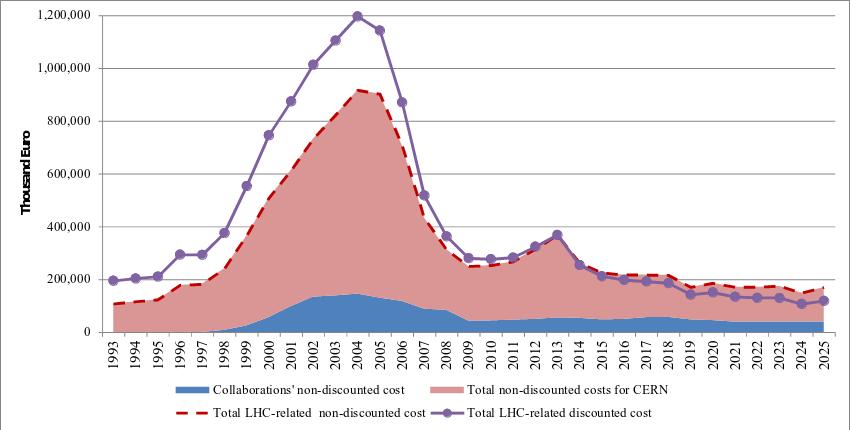}
 \end{center}
\vspace{-0.3cm}
\caption{\small \label{fig:costs} Time distribution of LHC costs
  (discounted and non-discounted).}
\end{figure}
Of the two components of the non-use value  $B_n$,  the quasi-option
value ${\rm QOV}_0$
includes serendipity effects, and it is thus intrinsically
uncertain~\cite{knight}. We thus take it as not measurable, we assume
that it is non-negative, and we set it to zero. 
The existence value ${\rm
   EXV}_0$ can be  proxied by willingness to pay (WTP). 
In environmental CBA, the existence value~\cite{pearce,EC}
is the benefit of preserving something known
to exist;  in our framework, it is the benefit of knowing that something
exists. 

In sum, our social accounting is
\begin{equation}\label{eq:npvres}
{\rm NPV}= \sum_i \frac{\left({\rm SC}_{t_i}+{\rm TE}_{t_i}+{\rm HC}_{t_i}+{\rm
   CU}_{t_i}\right)-\left({\rm K}_{t_i}+{\rm LS}_{t_i}+{\rm
    LO}_{t_i}+{\rm O}_{t_i}\right)}{(1+r)^{t_i}}+{\rm EXV}_0.
\end{equation}
Each variable in Eq.~(\ref{eq:npvres}) is split into
contributions  determined by other variables (e.g.,
scientists' salaries on the cost side, or additional profits of RI
suppliers on the benefit side), and it  is treated as
stochastic. 

We believe that this model is generally applicable to any RI, and its use
could help in the decision-making process. We 
now test it and validate it by applying it to the Large Hadron
Collider (LHC)~\cite{lhc}: arguably, the most stringent test of our
methodology. For each contribution on the r.h.s. of
Eq.~(\ref{eq:npvres}) we present our estimation of the corresponding
probability density (PDF), and use it to determine the PDF of ${\rm
  NPV}$ Eq.~(\ref{eq:npvres}).

{\it Costs}.  
LHC costs include  past and future capital and operational
cost born by CERN and the collaborations for building, upgrading and
operating the machine and experiments, including in-kind
contributions, for which there esists no integrated accounting. 
Three categories of costs have been
considered: i) construction capital costs, ii) phase 1 upgrade capital
costs, and iii) operating costs. 
CERN costs have been provided from the start up to 2025, while for 
collaborations we have reconstructed costs  using their own financial
reports, supplemente by our assumptions for years after 2013.
Integrated
past flows are capitalised and future costs discounted to 2013 Euro by
a 0.03 social discount rate (suggested for any infrastructure CBA in
Ref.~\cite{EC}), estimating  apportionment shares as needed.
The value of in-kind contributions has been estimated to be
$1.2\cdot 10^9$~\euro. We have reconstructed the  time distribution of
this total value over 1995 to 2008 (see Figure~\ref{fig:costs}), while
 CERN costs unrelated to LHC and costs for future upgrades have been excluded, as their benefits will occur
 beyond our time horizon.
Scientific staff costs have been assumed to balance the value of
 scientific output (see below) while for CERN administrative and
technical staff   we have assumed
 that 90\% of the cost   would have been borne regardless of the LHC. 
Our final estimate for the mean cost of the LHC is  
$\langle {\rm K}+{\rm LS}+{\rm
    LO}+{\rm O}\rangle= 13.5\cdot 10^{9}$~\euro.

\begin{figure}[t]
\begin{center}
  \includegraphics[width=0.6\textwidth]{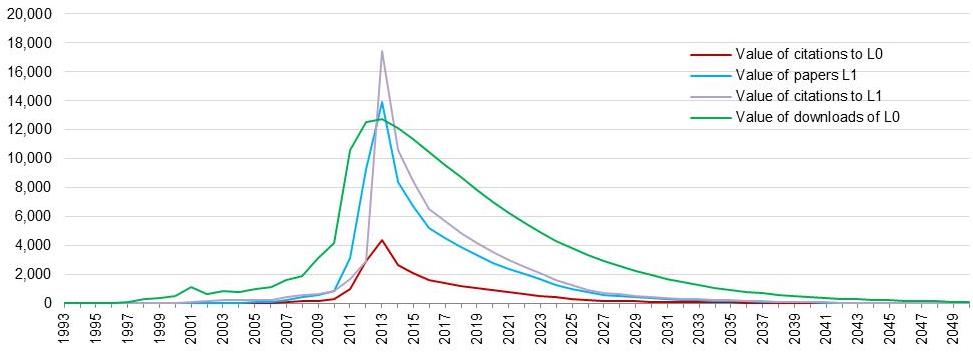}
 \end{center}
\vspace{-0.3cm}
\caption{\small \label{fig:knowledge} Economic value (constant k\euro
  2013) 
per year of citations to L$L_0$  and $L_1$ papers; value of $L_1$
papers; value of downloads of $L_0$  papers.}
\end{figure}
{\it Knowledge output.} The core benefit of the LHC to scientists
is publications. Publications produced by LHC
scientists ($L_0$)  have a value which is equal to their
production costs (scientific staff costs), hence neither
is included (see above). Benefits come from papers  ($L_1$)
by non-LHC scientists citing $L_0$ papers, with the benefit 
of papers 
citing these in turn  considered to be negligible.  
We proxy the MSV of $L_1$ papers through the average salary received
for time spent on doing research and writing. Our results,
based on an estimate of publication trajectories over a period of $N=50$
years starting with 2006 obtained 
through  a suitable 
model~\cite{bacchiocchi,carrazza,abt}, are summarized in Figure~\ref{fig:knowledge}: the
mean value of the corresponding benefits is $\langle {\rm SC}\rangle=280\cdot 10^{6}$~\euro.

\begin{figure}[t]
\begin{center}
  \includegraphics[width=0.6\textwidth]{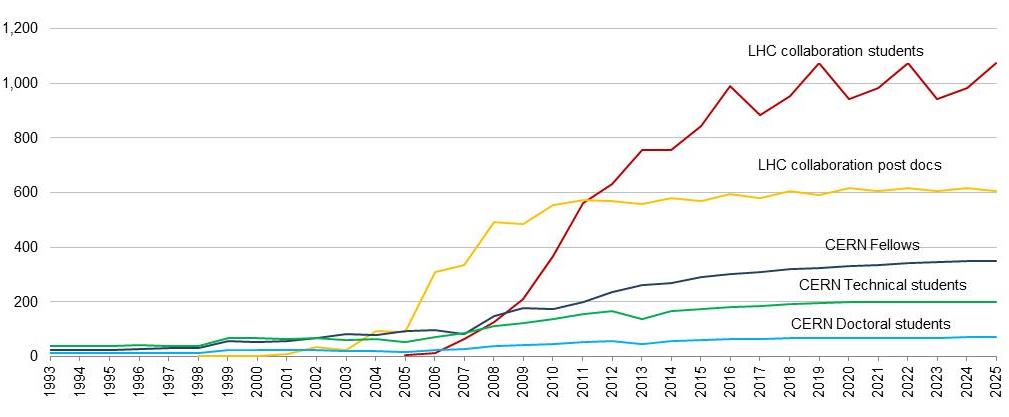}\\
  \includegraphics[width=0.3\textwidth]{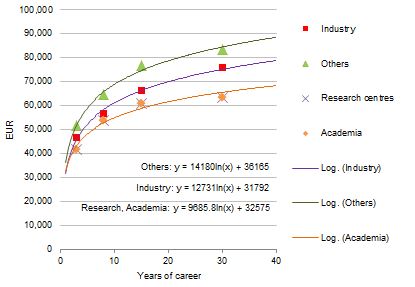}
  \includegraphics[width=0.3\textwidth]{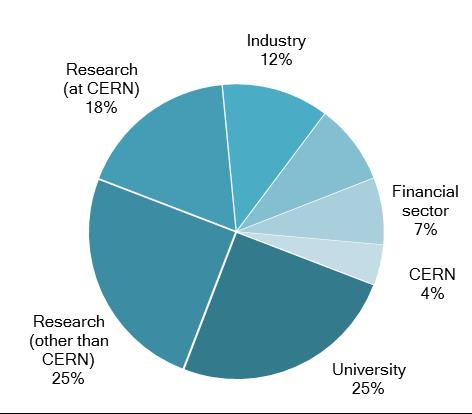}\\
  \includegraphics[width=0.3\textwidth]{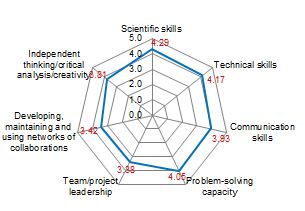}
  \includegraphics[width=0.3\textwidth]{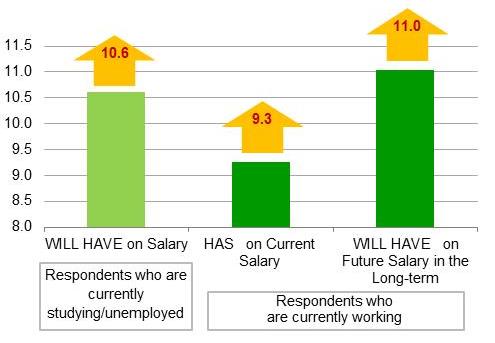}
 \end{center}
\vspace{-0.3cm}
\caption{\small \label{fig:human}  Top: Types and number of people
  benefitting from training at the LHC, historical data and
  forecasts. Center: Estimation of future average salaries (left);
  current employment sector of CERN alummni (right). Bottom:
 Perception of skill improvements  due to the LHC
  experience (left); percentage
  impact  on salary due to the LHC experience estimated by
  current students (light green)  and past-students (dark green) (right).}
\end{figure}
{\it Human capital.} The beneficiaries of human capital
formation~\cite{schopper,camporesi}  at
LHC over the time period  1993-2025 are 37000 young researchers:
19400 students   and 17000 post-docs.  (not
including  participants to schools or short trainings).
The  LHC benefit is valued as the PV of the
LHC-related incremental salary earned over the entire work
career (see Fig.~\ref{fig:human}). The
mean value of the corresponding benefits is $\langle HC\rangle= 5.5\cdot
10^{9}$~\euro.

\begin{figure}[h]
\begin{center}
  \includegraphics[width=0.8\textwidth]{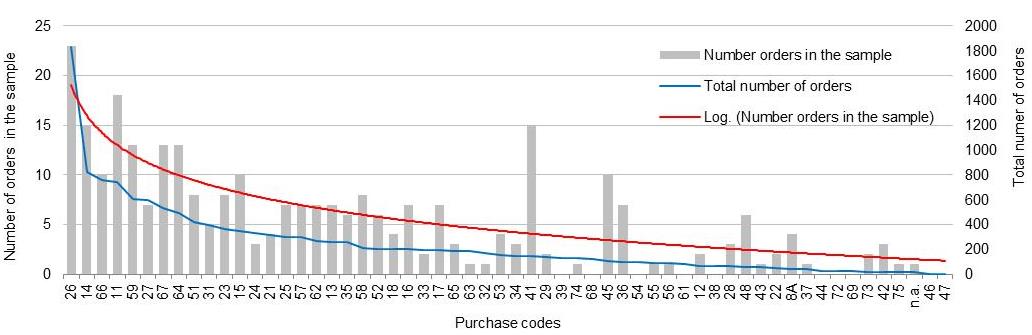}\\
  \includegraphics[width=0.4\textwidth]{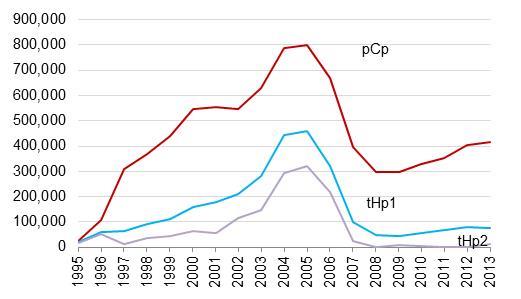}
  \includegraphics[width=0.4\textwidth]{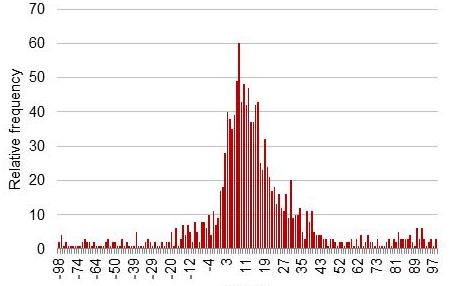}\\
  \includegraphics[width=0.4\textwidth]{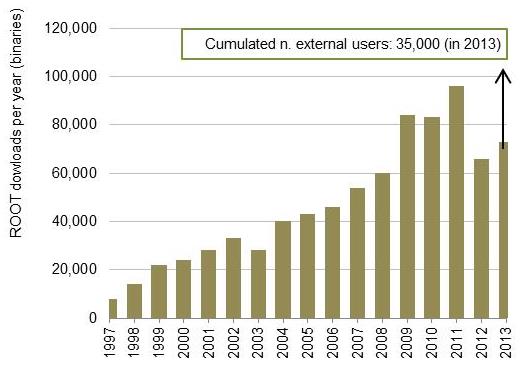}
  \includegraphics[width=0.4\textwidth]{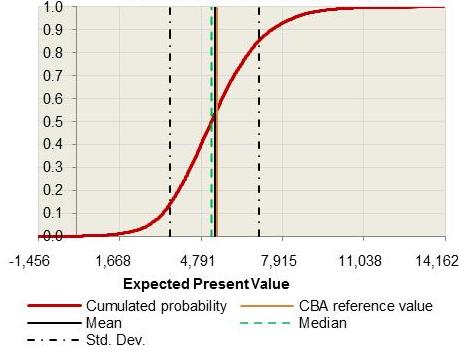}
 \end{center}
\vspace{-0.3cm}
\caption{\small \label{fig:tech} Top: Benefits to firms in the CERN supply
  chain from a sample of 300 orders by purchase code compared with all LHC
  orders {\tiny (CERN activity codes: 11 building work - 12
  roadworks - 13 installation and supply of pipes - 14 electrical
  installation work - 15 heating and air-conditioning equipment
  (supply and installation) - 16 hoisting gear - 17 water supply and
  treatment - 18 civil engineering and buildings - 21 switch gear and
  switchboards - 22 power transformers - 23 power cables and
  conductors - 24 control and communication cables - 25 power supplies
  and converters - 26 magnets - 27 measurement and regulation - 28
  electrical engineering - 29 electrical engineering components - 31
  active electronic components - 32 passive electronic components - 33
  electronic measuring instruments - 34 power supplies - transformers
  - 35 functional modules \& crates - 36 rf and microwave components
  and equipment - 37 circuit boards - 38 electronics - 39 electronic
  assembly and wiring work - 41 computers and work-stations - 42
  storage systems - 43 data-processing peripherals - 44 interfaces
  (see also 35 series) - 45 software - 46 consumables items for
  data-processing - 47 storage furniture (data-processing) - 48 data
  communication - 51 raw materials (supplies) - 52 machine tools,
  workshop and quality control equipment - 53 casting and moulding
  (manufacturing techniques) - 54 forging (manufacturing techniques) -
  55 boiler metal work (manufacturing techniques) - 56 sheet metal
  work (manufacturing techniques) - 57 general machining work - 58
  precision machining work - 59 specialised techniques - 61 vacuum
  pumps - 62 refrigeration equipment - 63 gas-handling equipment - 64
  storage and transport of cryogens - 65 measurement equipment (vacuum
  and low-temperature technology) - 66 low-temperature materials - 67
  vacuum components \& chambers - 68 low-temperature components - 69
  vacuum and low-temperature technology - 71 films and emulsions - 72
  scintillation counter components - 73 wire chamber elements - 74
  special detector components - 75 calorimeter elements – 8A radiation
  protection - n.a. not available)}.  Center: CERN external procurement  -
  commitment for total and high-tech orders {\tiny (pCp: Past CERN
  procurement - commitment (kEUR 2013) – tHp1: Total high-tech
  procurement - commitment (kEUR 2013) – tHp2: Total high-tech
  procurement - commitment - only orders >50 kCHF (k\euro
  2013))} (right); distribution of  EBITDA 2013 from ORBIS in firms at NACE
  industry levels matched with CERN codes (right). Bottom:  ROOT download
  data (left); ENPV Cumulative distribution function 
conditional to PDF of critical
  variables (k\euro 2013) (right).}

\end{figure}

{\it Technological spillovers.} Benefits to LHC-related supplier
firms consist of
incremental profits gained thanks to technology transfer and knowledge
acquired. We estimate these based on 
LHC-related procurement orders, 
categorised according to  high-tech
activity codes, which we forecast up to 2025, and then used to
determine incremental turnover for the suppliers through
estimates of economic utility/sales ratios from Ref.~\cite{bianchi,bianchi2},
(based on interviews to CERN suppliers) and
EBITDA data for companies in related sectors extracted from the
ORBIS database~\cite{orbis} (see Figure~\ref{fig:tech}).
Further benefits come from software developed for the LHC 
and made available for
free: 
ROOT (about 25000 users outside 
physics, mostly in the finance sector) and
GEANT4 (used  e.g. in medicine for simulating radiation damaging on
DNA), whose benefits are estimated  as the avoided cost for
the purchase of an equivalent commercial software (ROOT) or the cost
required for development of an analgous tool (GEANT4).
The
mean value of these benefits is $\langle TE\rangle=
5.4\cdot 10^{9}$~\euro. 
\clearpage

\begin{figure}[t]
\begin{center}\begin{minipage}{.48\linewidth}\begin{center}
  \includegraphics[width=0.6\textwidth]{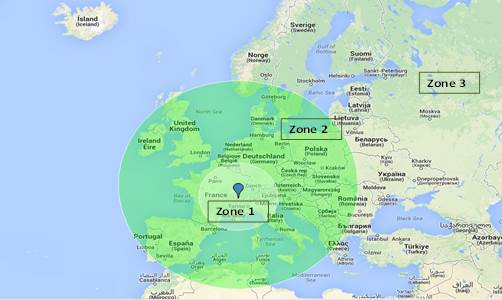}\\
  \includegraphics[width=0.6\textwidth]{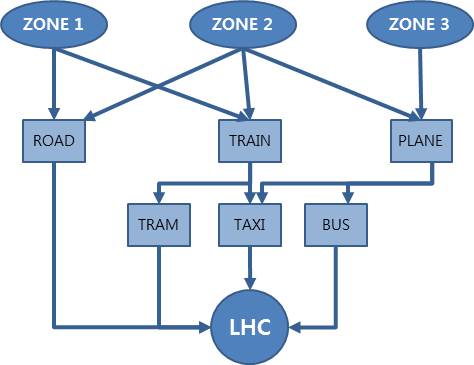}\\
  \includegraphics[width=0.6\textwidth]{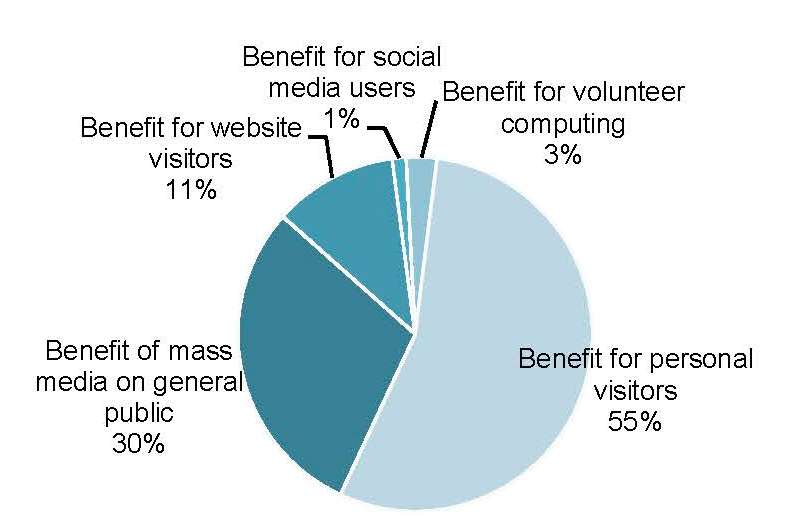}\\ 
\end{center}\end{minipage}\begin{minipage}{.48\linewidth}\begin{center}
 \includegraphics[width=0.6\textwidth]{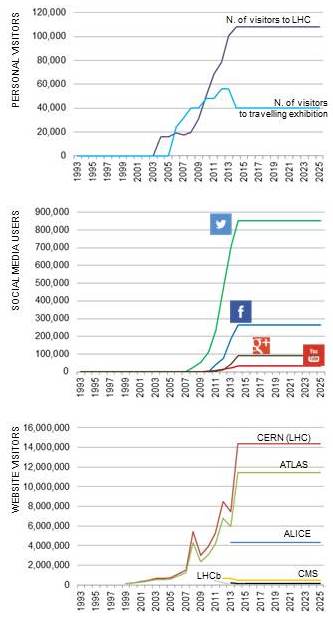}
\end{center}\end{minipage}  
 \end{center}
\vspace{-0.3cm}
\caption{\small \label{fig:cult} Left: (from top to bottom)
Travel zones for CERN  for visitors; CERN
visitors by mode of transport; share of benefits by type of
outreach activity (Cumulated impact to 2025).  Right: benefits to personal,
visitors,  social media users and website visitors.}
\end{figure}
{\it Cultural effects.}  These are benefits of LHC to the general public
visiting CERN, and taking advantage of  its exhibitions, websites, and 
outreach activities.  Benefits from on-site visitors
are determined using
the revealed
preference method~\cite{clawson}, with the
MSV of the time spent in travelling  obtained from
HEATCO~\cite{heatco} data  (see Figure~\ref{fig:cult}).
Further benefits come from LHC-related social media 
 and website visits, with
the MSV
of time of the general public  proxied by the hourly value of per
capita GDP (see Figure~\ref{fig:cult}). Finally, two 
CERN projects exploit computing time donated
from volunteers to run simulation of particle collisions, with
WTP revealed by time spent. 
The
mean value of cultural effects is $\langle CU\rangle=2.1\cdot 10^{9}$~\euro.

\begin{figure}[t]
\begin{center}
  \includegraphics[width=0.4\textwidth]{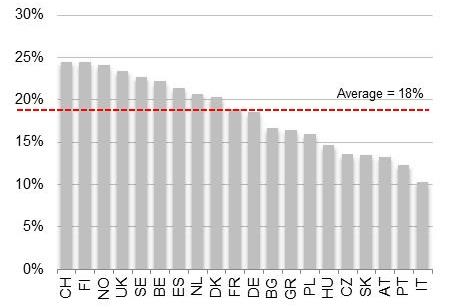}
  \includegraphics[width=0.4\textwidth]{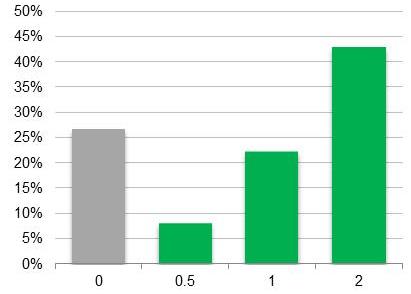}
 \end{center}
\vspace{-0.3cm}
\caption{\small \label{fig:noub} Share of adult population (18-74
  years old) with at least tertiary education (left); average annual
  WPT of the respondents to the survey (right).}
\end{figure}
{\it Non-use value.} A contingent valuation
approach (consistent with the NOAA 1993 protocol~\cite{noaa}) is used to
determine  social
preferences for the non-use value of the LHC as discovery device, a
public good with unknown practical use, proxied by WTP. 
Samples of students in four European countries were asked their WTP 
an annual fixed donation for 30 years; results were used to determine 
the WTP of people with tertiary education in CERN Member
States, and  people from non-Member States, based
on share of visitors (see Figure~\ref{fig:noub}). The mean non-use value is found
to be  $\langle {\rm EXV}_0\rangle= 3.2\cdot 10^{9}$~\euro.

\begin{figure}[t]
\begin{center}
  \includegraphics[width=0.4\textwidth]{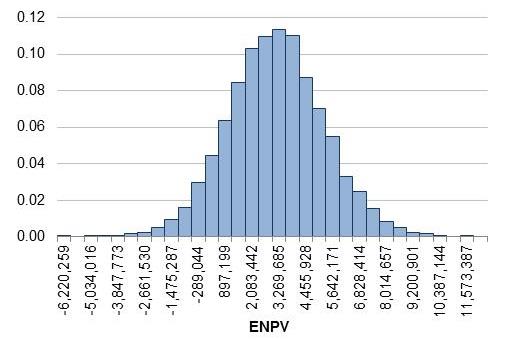}
  \includegraphics[width=0.4\textwidth]{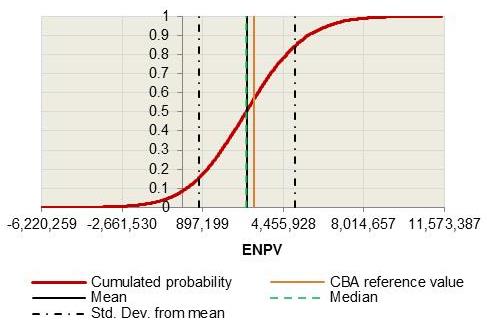}
 \end{center}
\vspace{-0.3cm}
\caption{\small \label{fig:enpv} Net present value  PDF (left) and
  cumulative distribution (right).}
\end{figure}
We have 
determined the PDF for the ${\rm NPV}$ Eq.~(\ref{eq:npvres})
by running a Monte Carlo simulation (10000 draws conditional to 19
stochastic variables)~\cite{eckardt,pouliken,salling}. The final PDF
and
cumulative probability distribution
for the  ${\rm NPV}$ are shown in Figure~\ref{fig:enpv}, with
a $3\sigma$
Monte Carlo error below 2\%. We find that the expected
${\rm NPV}$ of the LHC is around 2.9~billion~\euro, with a probability of a
negative ${\rm NPV}$ smaller than 9\%.  The expected Benefit/Cost ratio
is around 1.2 and the expected internal rate of return is 4.7\%. 

We have thus shown how a
social CBA probabilistic model can be applied to evaluate a large
scale research infrastructure, based on empirically feasible methods.
The unpredictable benefits of science (if any) are not included in our
analysis: they will remain as an extra bonus for future generations, donated to
them by current taxpayers.  

{\bf Acknowledgements:} We are grateful for comments on earlier
versions of the manuscript 
to Antonella Calvia G\"otz (EIB), Albert De Roeck (CERN), Andr\`es
Fai\~na (University a Coru\~na), Anna Giunta (University of Rome III),
Diana Hicks (Georgia Institute of Technology), Per-Olov Johansson
(Stockholm School of Economics), Mark Mawhinney (EIB/JASPERS), Giorgio
Rossi (ESFRI), Herwig Franz Schopper (CERN Director General Emeritus),
Florian Sonneman (CERN), Alessandro Sterlacchini (Universit\`a
Politecnica delle Marche), Anders Unnervik (CERN), Witold Wilak
(European Commission) and several others. 
This paper has been produced in the frame of the project
‘Cost-Benefit Analysis in the Research, Development and Innovation
Sector’ sponsored by the EIB University research programme
(EIBURS). The findings, intepretatations and conclusions presented in
the paper should not be attributed to the EIB or other institutions.

\clearpage

\begin{center}{\large\bf Supplementary material}\end{center}
{\it 1. Costs.} Capital and operational expenditures related to LHC
have been estimated as follows. Budgetary allocations from CERN to LHC
have been recovered from data communicated to us by the CERN Resource
Planning Department drawing from the CERN
Expenditure Tracking (CET) system (Account category, type, year,
program at 31 March 2014). These data cover all CERN program and
subprogram expenditures in current CHF, from January 1993 to 31
December 2013. The programs include: Accelerators, Administration,
Central Expenses, Infrastructure, Outreach, Pension Fund, Research and
Services. Cost for each Program are disaggregated in various
Subprograms (e.g. under Accelerators there are 19 Subprograms, such as
the SPS Complex, LHC, LEP, General R\&D, etc.). In turn each of these
items shows expenditures on materials, personnel, financial costs, and
others, broken down into 
recurrent and non recurrent expenditure. We have excluded financial
costs (such as bank charges and interests)  and we he
have identified the expenditure that can be attributed to the LHC. In
many cases it was necessary to estimate an apportionment share to LHC
of the expenditure for each item, which we have done
based on interviews with CERN
staff. Current CHF values have been first accounted in constant 
2013 CHF by considering the yearly change of average consumption prices from
IMF World Economic Outlook (October 2013), then expressed in \euro\ at
the exchange rate 1~CHF=0.812~\euro\ (European Central Bank,
average of daily rates for year 2013: {\tt\footnotesize
  http://www.ecb.europa.eu/stats/exchange/eurofxref/html/eurofxref-graph-chf.en.html}). 
The results is provided as a
supplementary table.  Past values have been capitalized to $t_0=2013$ with
a 0.03 social discount rate (EC, Guide to Cost-Benefit Analysis,
2014). Ten per cent of CERN Administration, Central expenses,
Administrative and Technical personnel have been attributed to LHC,
based on the hypothesis that, in a counterfactual history without LHC,
the CERN would have in any case sustained most of such costs, and
given the observation of past trends before the start-up of LHC
operations. A sensitivity analysis of the impact of taking a higher
share of such costs apportioned to LHC shows that the NPV remains
positive up to 75\% share attributed to LHC, without changing any
other hypothesis. Scientific personnel costs of CERN have been
identified from the reports “CERN Personnel Statistics”, available for
each year. The share of this part of the personnel every year is
between 19\% in 1993 and 32\% in 2013. This share of costs is assumed
to balance with the contribution of CERN scientists to the value of
the LHC publications, similarly to what we assume for non-CERN
scientists in the collaborations. To these direct CERN costs
we have  added  in-kind contributions from member
and non-member states. These are mainly equipment made available for
free to CERN by third parties and for which in Annual Accounts
(Financial Statements) 2008 (CERN/2840 CERN/FC/5337) 
a cumulative asset value of $1.47\cdot10^9$~CHF
 is recorded, combining in kind-contribution to the LHC
machine and the detectors. The attribution year by year of this
cumulative figure has been done assuming the same trend as CERN
procurement expenditures. The forecast for 2014-2025 of CERN expenditures has
been communicated to us by CERN staff, based on the ‘Draft Medium Term
Plan 2014’ (personal communication April 2, 2014). Again, we have
implemented an apportionment to LHC of each expenditure item. As all
values were given to us in constant CHF 2014, these were first
converted in CHF 2013, and then future values discounted to 2013 by
the 0.03 rate. As for personnel costs a share of 32\% of scientific
staff (corresponding to the actual share of scientific staff cost in
2013) was assumed constant for the future years, and deducted from
cost. We have not included any forecast of further in-kind
contribution.  
For the expenditures of the collaborations we have limited the
analysis to the four largest experiments (ATLAS, CMS, ALICE, LHCb); our main
sources have been the  Resource Coordinators of each
Collaboration. We have analysed the expenditure data particularly from
these sources: “CMS Summary of Expenditure for CMS Construction for
the Period from 1995 to 2008” (CERN-RBB-2009-032)  ; “CMS upgrade status report”
(CERN-RBB-2014-056); “Draft Budget for CMS Maintenance \& operations in the Year
2014” (CERN-RBB-2013-086); “Addendum No. 6 to the Memorandum of Understanding for
Collaboration in the construction of the CMS Detector” (CERN-RBB-2013-070/REV);
Addendum No. 7 to the Memorandum of Understanding for Collaboration in
the upgrade of the CMS Detector” (CERN-RBB-2013-127); Addendum No. 8 to the
Memorandum of Understanding for Collaboration in the Upgrade of the
CMS Detector” (CERN-RBB-2013-128); “Memorandum of Understanding for Maintenance
and Operation of the ATLAS Detector” (CERN-RBB-2002-035); “ATLAS Upgrade Status
Report 2013-2014” (CERN-RBB-2014-022); “Request for 2014 ATLAS M\&O Budget”
(CERN-RBB-2013-079); “Memorandum of Understanding for Maintenance and Operation
of the LHCb Detector” (CERN-RRB-2002-032.rev-2008); “Addendum No. 01 to the Memorandum of
Understanding for the Collaboration in the Construction of the LHCb
detector” (CERN/RBB 2012-119A.rev-2014); “Status of the LHCb upgrade” (CERN-RRB-2014-033); “RRB
Apr.2014” (CERN-RRB-2014-039); for ALICE data, the source is a
personal communication (7 May 2014) comprising
data such as Core Expenditure 2007-2013, Construction costs, including
Common Fund, per system, M\&O A-budget and B-budget. 
Fifteen more reports have been processed by us for the analysis of costs
(detailed list available with the authors upon request). Forecast of
future expenditures of the collaborations have been based on the same
sources. When only cumulative data at a certain year were available,
appropriate hypotheses about the yearly distribution have been
made. For the LHCb Collaboration some missing yearly data have been
assumed. We have not considered the cost implications of the High
Luminosity Project and of the LHC Upgrade Phase 2 as they mostly will
run after out time horizon. To avoid double counting, the CERN
contribution to the collaborations have been excluded by their
expenditures. As for CERN, the scientific personnel cost of the
collaborations (paid by their respective Institutes) has been taken as
balancing the value of the scientific publications, and excluded from
the grand total of cost. The overall trend of CERN
LHC-related and collaborations expenditures was shown in Fig.~1. 
While the information up
to 2013 has been taken as deterministic, the forecast 2014-2025 has
been treated as stochastic. A normal distribution of the total cost
has been assumed with mean equal to $1.966\cdot 10^9~\euro$, and a standard
deviation compatible with mean $\pm50\%$ as asymptotic values, based
on interviews on the most optimistic and pessimistic future
scenarios. We have not included decommissioning costs as we have no
information on them. For the same reason we have also not tried to
forecast accidents or negative externalities. 
\begin{figure}[t]
\begin{center}
  \includegraphics[width=0.8\textwidth,page=1]{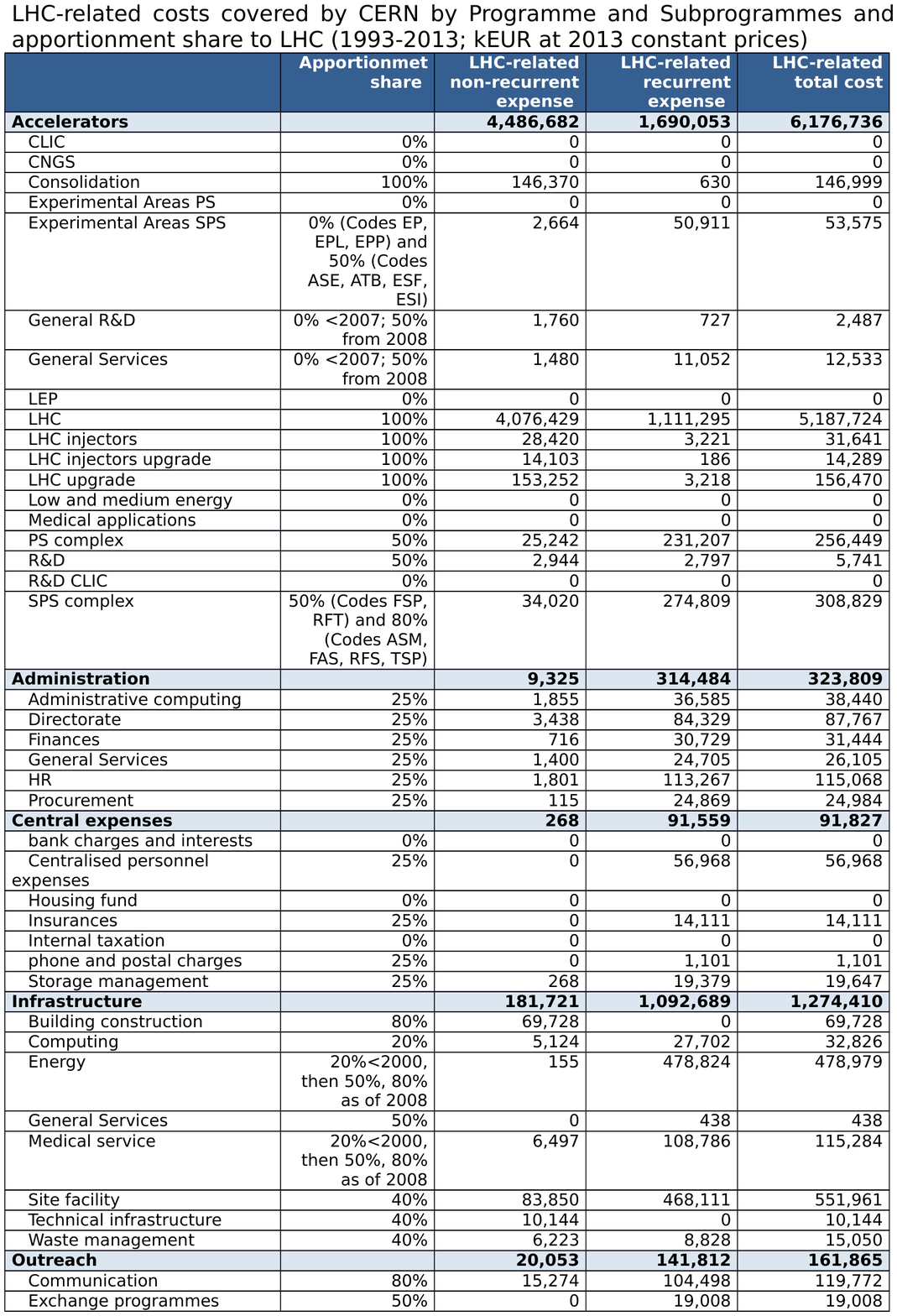}
\vskip-3.cm

  \includegraphics[width=0.8\textwidth,page=2]{supptable.pdf}
\end{center}
\end{figure}

\clearpage

{\it 2. Value of Publications.} The past (1993-2012) number of
LHC-related scientific publications $L_0$ (including CERN and
collaborations) has  been extracted from the inSPIRE database
({\tt http://inspirehep.net/}) by Carrazza, Ferrara and Salini [17], as
part of this project. The data include published articles and
preprints. Citations to these up to 2012 have been
retrieved from the same source. In order to forecast the number $EL_0$
of $L_0$
publications 2013-2025 we have applied a double exponential model of
the form [16,17]
$$  EL_0(t)=\alpha_1 \alpha_2 \exp\left[-\beta_1(T-t)\right]
\left[1-\exp\left[-\beta_2(T-t)\right]\right],$$
with $\alpha_1=65000$, $α_2=2$ $\beta_11=0.18$, 
$\beta_2=0.008$, $T=50$, $t=2006$. 
The forecast of the number of $L_1$ publications over the
years 2013-2050 has been based on observed pattern of average number
of citations per paper, without assuming any new spike after the one
related to the discovery of Higgs boson. We have also estimated the
citations to $L_1$ papers by $L_2$ papers. Again, the number of $L_2$ papers
until 2012 is based on inSPIRE, while to forecast 2013-2050 we
assume 4 citations per paper, in line with the previous years. To
these figure we have added downloads, which for the field of High
Energy Physics are available from arXiv ({\tt arxiv.org}), which we
used  for 1994-2013, while in order to  forecast until 2050 we have assumed
the same average in future as the past (64 downloads per paper). This
average number of downloads has been applied to $L_0$ papers. 
The benefits are thus: the value of $L_1$ papers; the value of $L_1$
citations to $L_0$ papers; the value of $L_2$ citations to $L_1$
papers.  The
value of $L_0$ papers cancels against its cost and it is not
included. The value of $L_2$ papers and beyond,  and citations to them, is
considered to be
be negligible.
All values are discounted at the 3\% social discount rate. 
After this baseline estimation, risk analysis has been performed on
the total present value of the publications, assuming a mean of
$277\cdot 10^6~\euro$ 
and a standard deviation of  $103\cdot 10^9~\euro$.  These parameters in turn
are based on a Montecarlo simulation (10,000 draws) of a range of
values for the following variables: number of references to $L_0$ papers
in papers $L_1$ (Ref.~[18]); percentage of time of
scientists devoted to research (based on interviews to LHC users);
papers produced per year per scientist (interviews); average salary on
non-LHC scientists ({\tt payscale.com}); time per download (interviews);
time per citation (interviews).

{\it 3. Human Capital}. We have considered five types of students or young
researchers: CERN doctoral students; CERN technical students; CERN
fellows; users under 30 years; users between 30 and 35 years. The
source of data are the yearly reports ‘CERN Personnel statistics’ from
1995 until 2013. Number of incoming students year by year for each
type and average stay have been estimated, based on interviews with CERN
Human Resources Department staff. Future incoming student flows
have been extrapolated from past trends. Specific apportionment of these
flows to LHC have been computed with the following estimates: 30\%
(for the period 1993-1998); 50\% (1999-2001); 70\% (2002-2007); 85\%
(2008-2025). Specific additional assumptions have been made for each
of the five types in order to derive the flow of annual incoming
students over the years 1993-2025. The estimated total cumulated
figure for students and young researchers is 36771 to 2025. In order to
estimate the economic benefit to each of these, a survey, directed to
students
and former students, 
was performed between May and October 2014
and in March 2015, both through an on-line questionnaire and direct
interviews at CERN. Information from 384 interviewees coming from 52
different countries has been collected: 75\% of respondents are male;
38\% are 20-29 years old and 43\% are 30-39 years old; 65\% of
respondents are related to the CMS Collaboration and 22\% to ATLAS. The
survey gives us an ex-ante or ex-post perceived LHC premium on
salary. As the two averages are very similar, we have considered more
reliable the
ex-post data,  i.e. the premium declared by former
students who have already found a job: it is  equal to 9.3\%. This percentage
premium has been applied to the average annual salary at different
experience levels, retrieved from the payscale database
({\tt www.payscale.com}). In
particular, we have classified salaries by experience level (entry,
mid-career, experienced, and late career) for different jobs in the
USA {see e.g. {\tt
    http://www.payscale.com/research/US/Job=Electronics\_Engineer/Salary}), 
grouped in four broad sectors: industry, research centres,
academia, others (the latter including for instance finance, computing
and civil service). A distribution of the number of CERN
students across these broad sectors has been assumed: for CERN
technical students we have assumed that only 10\% will go either in
research centres or in the academia, and 45\% respectively in the other
two sectors; for the other students we have assumed a destination in
research and academia for 60\%, and  20\% each for the others. The four
aforementioned career points have been interpolated with a
logarithmic function. Given the average salary in each sector, the
premium declared by interviewees, and the assumed shares of students
finding a job in each sector, we have computed this component of the
human capital effect. Considering that the difference between the pay
in research and academia and the two other sectors combined is between
13\% and 18\% (increasing with the level of experience), and that 14\% of
the former students who have participated in the survey have been diverted
to better-paid jobs in  industry of other sectors, an additional
premium between 2-3\% (triangular PDF with average and mode both equal
to 2.5\%) has
been applied. The total 11.8\% premium has been attributed to each
student over a career spanning 40 years, with the implication, for
example, that the cohort of 2025 student will enjoy the benefit up to
2065. The total number of students has been taken as a triangular PDF
with maximum and minimum equal to  $\pm 15\%$ of the mode and mean. 
All values are
discounted, which, because of the long time span, 
roughly halves the cumulated benefit in comparison
to its undiscounted value.  

{\it 3. Technological spillovers.} The total value of CERN procurement
by year and by activity code has been recovered from the CERN
Procurement and Industrial Services Companies (personal communication,
October 2013). A random sample of 300 orders exceeding
$10^5$~CHF in nominal value  has been extracted from a
data set in turn extracted for us by the aforementioned CERN office. 
Each order has
been classified with the help of expert CERN staff according to a
five-point 
scale: 1) very likely to be off-the-shelf orders with low
technological intensity; 2) off-the-shelf orders with an average
technological intensity; 3) mostly off-the-shelf orders by usually
high-tech and requiring some careful specifications; 4) high-tech
orders with a moderate to high specification activity intensity to
customize product for LHC; 5) products at the frontier of technology
with an intensive customization work and co-design involving CERN
staff. An average technological intensity has been attribute to each
CERN activity code, and  we have considered as high-tech the codes with
average technological intensity class equal or greater than 3. This
has led to the identification of 23 activity codes. 
Procurement value has then been
computed only for these codes, and turned out to be  35\% of the total
of procurement expenditures. This would be only 17\% if we exclude orders below
$5\cdot 10^4$~CHF, and symmetrically 58\% if we include orders below this
threshold and for other activity codes. We took a triangular
distribution with average and mode model equal to 
35\% and minimum and maximum as
above. A share of 84\% of yearly total expenditures of collaborations
is attributed to external procurement, using the same share as
CERN. This share has been used also for the future forecasts of both
the CERN and collaborations up to 2025, based on the previous forecast
of cost trends. For the collaborations, which are known to include a
significantly higher share of high-tech orders, we assume a triangular
distribution of the share of high-tech procurement with
average and mode equal to 58\%, minimum set to 40\%, and  maximum to
75\%.  We have then identified 1,480 firms from the ORBIS database
[23] in the year
2013 and in six countries (Italy, France, Germany, Switzerland, UK,
USA), selected because they received 78\% of the total  CERN
procurement
 expenditure between 1995 and 2013 (data on procurement commitment
by country provided by CERN staff, October 2013). In selecting this
sample, we have considered companies whose primary activity matches
with the corresponding CERN activity codes. The following NACE
sectorial codes have been considered: manufacture of basic metals (24);
manufacturing of structural metal products (25.1); forging, pressing,
stamping and roll-forming of metal (25.5); manufacturing of other fabricated
metal products (25.9);  manufacturing of computer, electronic and optical
products (26); manufacturing of electrical equipment (27); manufacturing of
machinery and equipment not classified elsewhere (28); specialised construction
activities (43); telecommunications (61); computer programming,
consultancy and related activities (62); information service
activities (63). After having observed the EBITDA margin sample
distribution, we have computed an average (13.1\%) and standard
deviation EBITDA weighted by country, and used these parameters to
define a normal distribution of the EBITDA. We have then estimated the
incremental turnover over 5 years by the LEP average utility/sales
ratio to be equal to three, based on the results of Refs.~[21,22].
 Based on these sources we assumed a triangular
distribution with mode equal to the mean, 
minimum 1.4, maximum 4.2. This ratio
has been applied to the high-tech procurement of both CERN and
collaborations. We have finally computed the additional
sales times EBITDA margin, thus estimating the additional profits of firms in
the LHC supply chain. All the detailed data are available upon
request.  

Out of several open-source software codes  available from CERN we have
identified ROOT and GEANT4 as mostly developed in relation to
LHC computing needs. Non-CERN ROOT users  outside the
high-energy physics community 
are estimated to be about 25000  worldwide in
2013, in addition to about 10000 HEP users, on
the basis of yearly download statistics of the software code ({\tt
  https://root.cern.ch/drupal/content/download-statistics})   as
well as interviews and personal communication with CERN Physics
Department staff. We then determined future trends  
based on estimates of CERN staff on the basis 
of past yearly downloads, which are  55000 in
2025. This has been taken as a stochastic variable with a triangular distribution
and a range of $\pm 20\%$ about equal average  and mode. 
The number of new users
by year has been estimated based on interviews to CERN staff. The market
prices of several comparable  commercial software codes has been
analyzed, based on interviews to CERN staff. The 
range of avoided costs, depending on
computing needs, goes from zero (if the
 R open-source statistical analysis code was used instead) to 17000 Euro per year for a one-year license (if
Oracle advance analytics was used). We have assumed a triangular yearly cost-saving
PDF for each ROOT user, with average and mode equal to 1500~\euro, 
minimum set to 1000~\euro, and
maximum 2000~\euro. Based on interviews, we have assumed a trapezoidal
PDF for the number of usage years, with modes equal to  3 and 10;
minimum 0; maximum
20. Then number of users, multiplied by the 
avoided cost per year is then discounted and
summed to compute the PV of the ROOT-related benefit. 
For GEANT4 ({\tt http://geant4.web.cern.ch/geant4/license/}) we have
identified about fifty research centres, space agencies and firms in
which it is routinely used (not including hospitals  which use GEANT4 for
medical applications).
 Out of these we have made a distinction  between the 38 centres who
contributed in some form to the development of the code, and the
remaining ones.  The avoided cost is based on the production cost
of GEANT4 (around $35\cdot 10^6~\euro$ up to 2013, provided by CERN staff
and generated using SLOCcount {\tt www.dwheeler.com/sloccount}); the total CERN
contribution to this cost is estimated to be 50\%. The avoided cost 
for the aforementioned  38 centres is reduced to the contribution they
actually provided (assumed to be the same for each centre, thus 50\%
of 35 million euro divided by 38), while it is the full GEANT4 cost
for the remaining ones. A forecast to 2025 and a yearly avoided cost
has been then estimated. The total cumulated avoided cost has been
taken as symmetric triangular PDF $\pm30\%$ about a mode and mean both
equal to $2.8\cdot 10^9~\euro$.

{\it 5.Cultural effects.} The benefits and population variables
considered are: (1) number of on-site CERN visitors; (2) number of visitors to
CERN traveling exhibitions; (3) number of people reached by media
reporting LHC-related news; (4) visitors to CERN and collaborations
websites; (5) number of users of LHC-related social media (YouTube;
Twitter; Facebook; Google+); (6) number of participants in  two
volunteer computing programs.  
Data for (1) have been provided by the Communication Groups of CERN
and each collaboration from 2004 to 2013. The forecast to 2025 (here and
for the other variables) has been assumed to be given by
a constant yearly value,
equal to the average of the last years. We have assumed 80\% of
overlap between visitors to experiment facilities and the permanent
CERN Exhibitions (‘Microcosm’ and ‘Universe of Particles’ in the Globe
of Science and Innovation); moreover, only 80\% of visitors to CERN
have been attributed to the LHC. The valuation of the benefit is based
on the segmentation of visitors in three  areas of origin
with increasing
distance from CERN (see Figure~5), and by average travel costs for each zone, based
on seven origin cities taken as cost benchmarks. For each zone a
transport mode combination and length of stay have been assumed (see
Figure~5). The 
three zones and the share of visitors for each zone are based on data
provided by the CERN Communication Group (personal communication from
October 2013 onward); additional costs have been
estimated including accommodation and meals (data extracted from the
CERN website).  The value of travelers' time is based on HEATCO [25]
for each member state and for some non-members. Based on the
distribution of visitors by country and mode of transportation, 
we have estimated an overall
distribution of visitors based on the following assumptions: trapezoid
distribution for air travelers (minimum equal to 5; maximum
equal to 45,  first mode equal to  22 and second  equal to
27, all in \euro/hour); triangular distribution for travel by car and train 
(mean and mode equal to 18; minimum 6 and  maximum 30).  
For variable (2), we have used the number of past visitors as provided by
CERN (between 30000 and 70000 for the cumulative period 2006-2013). We
have assumed a constant number of 
40000 visitors per year during from 2014 to 2035. 
The WTP is prudentially assumed to be just 1~\euro\ 
per visitor (assuming local transport).  
For the variable (3) we have conservatively considered only the news
spikes in September 10th 2008 (first run of LHC) and July 4th 2012
(announcement of discovery of the Higgs boson). Sources for these
point estimates are: New Scientist (2008) and
{\tt
  http://cds.cern.ch/journal/CERNBulletin/2012/30/News\%20Articles/1462248}.   
We have assumed, based on interviews, that the time devoted to LHC
news per head is 2 minutes. We have treated the audience as a
stochastic variable,
assuming a triangular distribution (minimum zero, maximum one billion,
average and mode equal to 5000 million). 
The value of time of the target audience has been estimated
based on current GDP per capita in the average CERN Member
States and the USA (for 2013, using IMF data), and the number of working days
per year (8 hours times 225 working days). This is treated as a stochastic,
triangular distribution, with minimum equal to 3~\euro; maximum
42~\euro, and mode and mean equal to 17~\euro. 
Website visitors (4) have been determined  on the basis 
of historical data on hits until
2013-2104 (source CERN and
collaborations’ Communication Groups). Our forecast is conservatively
based assuming that the value at  the last available observation
remains constant. The benefit comes from 
the number of
minutes per hit from users of the websites, estimated to be a triangular
distribution with average and mode equal to  minutes, and ranging from
0 to 4 minutes.  
For social media (5), we used data provided  by CERN and
collaborations,
attributing to the LHC  80\% of the hits to CERN-related
social media and 100\% of those related to the collaborations. We
used historical data until 2014 and for the subsequent years we
have taken the last year data as constant. The average stay time is
assumed for all social media to be distributed according to a triangular
distribution with average and mode equal to 0.5 minutes per capita,
ranging from zero to one minute. Time is then valued as  above.  
Volunteer computing (6) is represented by  two LHC-related programs:
SIXTRACK and TEST4THEORY. The stock number of volunteers in 2013 has
been provided  by the CERN PH Department (via personal
communication);  based on this we have assumed a rate of increase
from the start years (respectively 2007 and 2001). A forecast of
the stock has been given to us to 2025 by the same source, and again
we have assumed a yearly rate of change over the years 2014-2025. The
opportunity cost is the time to download, install, and configure the
programs (15 minutes per capita una tantum) and the time spent in
forum discussions (15 minutes per month per capita). Again, time is
valued as above.

{\it 7. Existence value of discovery.} The survey on WTP for the LHC as a
public good has been performed in Milan in October-November 2014, and
in Exeter (UK), Paris (France), A Coru\~na (Spain) in February-March
2015: 1027 questionnaires have been collected. The average time spent
answering the questionnaire (28 questions available upon
request) was about 25 minutes. The respondent was first given a one page
summary of the LHC Wikipedia page as an information set. Geographical
distribution of respondents was 40\% from Italy, and 20\% each from Spain,
France and UK. Out of the total number of respondents, 85\% were
19-25 years old, 57\% were females, 64\% were in the humanities and
social sciences. Questions included: household composition, family
income, personal income, high-school background, previous knowledge of
research infrastructures, source of information, if any, on the LHC
and the Higgs boson discovery, whether the respondent has ever visited
CERN, interest in science, willingness to pay for LHC research
activities an economic contribution lump sum or yearly over 30 years,
in pre-set discrete amounts (zero, 0.5, 1, 2 Euro). We have then taken
the sample average yearly WTP, weighted by the number of respondents
by country, for  respondents who declared a positive annual WTP
(73\% of the total). This has given us  a sample distribution with three
discrete values (0.5, 1 and 2~\euro), and mode and maximum equal to 2. 
Each
annual WTP has then been multiplied (undiscounted) by 30 years. This per
capita WTP has been applied to 73\% of  18-74 year olds with at least
tertiary education  coming from  CERN Member
States (determined from Eurostat data 2013).  We have then added to
the previous target population 
an additional 21\% from CERN non-member states,
reflecting the share of personal visitors to CERN from non-member states
(visitor statistics provided by CERN staff as a personal
communication). We have treated  the per capita WTP as a stochastic
variable,  assuming a truncated triangular
probability distribution with maximum and mode equal to 2~\euro and
minimum equal to 0.1~\euro,
reflecting the sample distribution for non-zero values.

\end{document}